\documentclass[final,11p,times,twocolumn]{elsarticle}
\usepackage{soul}
\journal{Physics Letters B}

\usepackage{amsmath}
\usepackage{color}
\usepackage{graphicx}
\usepackage{hyperref}
\hypersetup{colorlinks, citecolor=blue, linkcolor=black, urlcolor=blue}
\usepackage[caption=false]{subfig}
\usepackage[american]{babel}
\usepackage{microtype}
 \usepackage[table,xcdraw]{xcolor}
\usepackage{color}
\usepackage{fancybox}
\usepackage{shadow}
\usepackage{graphicx}
\usepackage{pdfpages}
\usepackage{float}
\usepackage{fancyhdr}
\usepackage[margin=1in]{geometry}
\usepackage[none]{hyphenat}
\definecolor{darkblue}{rgb}{0.0,0.0,0.6}

\usepackage[normalem]{ulem}

\usepackage{geometry}
\usepackage{fullpage}
\begin{document}

\title{\sc{Constraining Anisotropic Universe Through Big Bang Nucleosynthesis:\\ A Case Study of The Bianchi Type-I Universe}}
%%%%%%%%%
\author[inst1]{Jiwon Park}
\ead{cosmosapjw@soongsil.ac.kr}

\author[inst2] {Sourav Mridha}
\ead{smridha03j@gmail.com }

\author[c,d,e]{Dukjae Jang \corref{cor1}}
\cortext[cor1]{Corresponding author} \ead{djphysics@gachon.ac.kr}

\author[inst2]{Mayukh R. Gangopadhyay \corref{cor2}}
\cortext[cor2]{Corresponding author}
\ead{mayukhraj@gmail.com}

\author[inst1]{Myung-Ki Cheoun}
\ead{cheoun@ssu.ac.kr}

%%%%%%%%%
\affiliation[inst1]{
        organization={Department of Physics and OMEG Institute, Soongsil University}, 
        state={Seoul}, 
        postcode={06978}, 
        country={Republic of Korea}
}

\affiliation[inst2]{
       organization={Centre For Cosmology and Science Popularization (CCSP), SGT University}, 
       city={Gurugram, Delhi- NCR}, 
       state={Haryana}, 
       postcode={122505}, 
       country={India}
       }
       
\affiliation[c]{
         organization = {Department of Physics, Gachon University}, 
         state={Gyeonggi-do}, 
         postcode={13120}, 
         country={Republic of Korea},
}

\affiliation[d]{
         organization={School of Liberal Arts, Korea University of Technology and Education}, 
         state={Chungcheongnam-do},
         postcode={31253}, 
         country={Republic of Korea}
}

\affiliation[e]{
      organization={Department of Physics, Incheon National University}, 
      state={Incheon}, 
      postcode={22012}, 
      country={Republic of Korea}
}

\begin{abstract}
The isotropy and homogeneity of our Universe are the cardinal principles of modern cosmology built on the definition of metric through the prescription by  Friedmann–Lema$\hat{i}$tre–Robertson–Walker (FLRW). From the aspects of geometry, the presence of anisotropy, inhomogeneity, or both are allowed in the metrics defined as the Bianchi type I and V metrics. In this letter, the Big Bang Nucleosynthesis (BBN) formalism, and the latest observational constraints on nuclear abundances are being used to put bounds on the global anisotropy offered in the Bianchi type I metrics, providing a new path to explore in the background of global anisotropy.
\end{abstract}
\maketitle
\section{Introduction}
%%%%%%%%%%%%%%%%%%%%%%%%%%%%%%%%%%%%%%%%%%%%%%%%%%%%%%%%%%%%
The cardinal principle on which modern cosmology developed, is based on the assumption of statistical isotropy and homogeneity of our Universe on larger scales, which is coined as the \textit{``Cosmological Principle"} or sometimes the \textit{``Copernican Principle"}. With the tremendous advancement of technology, observational cosmology has made it possible for us to treat cosmology as a precision science. On that note, many efforts have shown that the scale at which we expect anisotropy and homogeneity increases with better observations. Thus, one should check the presence of global anisotropy if that can be offered from the first principle.

Interestingly, mathematics offers at least two different classifications of metric which follow the continuum-sized Lie algebra on a Riemannian manifold and can provide global anisotropy in one case and in another both anisotropy and inhomogeneity at the global scale. These two metrics are named after the proposer: Luigi Bianchi, dubbed as Bianchi type I (B-I) and type V (B-V) metrics, respectively. The choice of Friedmann–Lema$\hat{i}$tre–Robertson–Walker metric (FLRW metric) was from the point of view of keeping the Copernican principle correct. But, other than that, from the mathematical point of view, the metric being B-I or B-V will not make any difference. Thus, naturally, if one wants to introduce global anisotropy or inhomogeneity, or both, the choice of B-I or B-V makes sense. Early interesting work on anisotropic universes can be found in \cite{Lemaitre:1933gd, Jacobs:1968zz}, and further recent developments in this regard can be found in (\cite{Caderni:1979ek}- \cite{Hertzberg:2024uqy}).

From the early Universe cosmology point of view, the abundance of primordial nuclides proposed by the standard Big Bang Nucleosynthesis (BBN) provides the most stringent constraint on any deviation from the standard model of evolution based on the FLRW metric for reference one can check (\cite{Jang:2016rpi}-\cite{Mathews:2018qei}).
Thus, it is natural to check the effect of global anisotropy or inhomogeneity on the abundance of primordial nuclides, in return providing a tight constraint on them. Thus, in this letter, we first propose to constrain the global anisotropy due to the B-I metric using the demands of BBN.

The rest of the paper is organized as follows:
In section~\ref{B1}, we discuss the basics of the Bianchi Type-I anisotropic Universe and introduce the evolution equation dubbed as the Friedmann equation in this scenario. In the section~\ref{bbn}, we discuss the constraints on global anisotropy, after performing the BBN analysis. Finally, we conclude this letter in section~\ref{conc} with our findings and constraints on the global anisotropy using the latest light element abundances. 

Finally, in this paper, we adopt a natural unit, i.e., $\hbar=c=k_B= 1$.

%%%%%%%%%%%%%%%%%%%%%%%%%%%%%%%%%%%%%%%%%%%%%%%%%%%%%%%%%%%%%%%%%%%%%%%%%%%%
\section{Bianchi Type- I Anisotropic Universe:}
\label{B1} % Ensure no problematic commands are used in hyperref-related contexts
In the B-I universe, the spacetime interval is determined by the following metric tensor:

\begin{eqnarray}
g_{\mu \nu} = \mathrm{diag}[-1, e^{2\beta_1(t)}a^2(t), e^{2\beta_2(t)}a^2(t), e^{2\beta_3(t)}a^2(t)],
\end{eqnarray}

where ${ a(t) }$ denotes the scale factor, and ${ \beta_i(t) }$ represents the directional anisotropies for ${ i=1,2,3 }$ in a (3+1)-dimensional spacetime. These functions satisfy the constraint of $\beta_1(t) + \beta_2(t) + \beta_3(t) = 0$, which ensures volume conservation despite anisotropic expansion. This condition is consistent with the theoretical framework based on a continuum Lie algebra over a Riemannian manifold.

In this B-I spacetime, assuming axial symmetry along the ${ i=3 }$ direction allows one to set $\beta_1(t) = \beta_2(t)$. Under this assumption, the shear stress tensor resulting from global anisotropy is given by,
\begin{eqnarray}
    \sigma_{a}^{\ b} = 
    \begin{pmatrix}
    \sigma    &      0     &    0  \\
       0      &   \sigma   &    0  \\
       0      &      0     &   -2\sigma
    \end{pmatrix},
\label{shear}    
\end{eqnarray}
and the anisotropic temperature (precisely, the temperature quadrupole deviation, the only one multipole deviation allowed in B-I) can be written as follows:
\begin{eqnarray}
T_a^{\ b} = \begin{pmatrix}
T_q   &   0    & 0  \\
0   &   T_q  & 0    \\
0   &   0    & -2 T_q
\end{pmatrix},
\label{Tab}
\end{eqnarray}
where $\sigma = \dot{\beta}_1(t) = \dot{\beta}_2(t)$, with the dot symbol denoting a time derivative, and $T_q \equiv \Delta T / T$ representing the relative temperature fluctuation. For simplicity, we assume that $\Delta T$ remains constant.

For the energy-momentum tensor, we adopt the general form:
\begin{eqnarray}
{\mathcal T}_{ab} \equiv \int p_a p_b f(E,T_{\mathrm{tot}})~{\mathrm d}P,
\label{emdef}
\end{eqnarray}
where $p_a$ is the four-momentum, $f(E,T_{\rm tot})$ is the particle distribution function, and $\mathrm{d}P$ denotes the Lorentz-invariant phase-space volume element. The particle distribution function $f(E, T_{\rm tot})$ depends on the particle species. We note that relativistic species dominate the early universe.

To determine the energy density and pressure of the energy-momentum tensor, in the presence of anisotropy, we adopt the total effective temperature $T_{\rm tot}$ within quadrupole $l=2$ expansion for $T_{ab}$: 
\begin{eqnarray}
T_{\rm tot} = T \left( 1+ T_{ab} e^a e^b \right),
\label{T_tot}
\end{eqnarray}
where $T$ is the typical temperature, $e^a$ is the unit vector in the direction of the three-momentum.

With the given energy-momentum tensor and the B-I metric tensor, one can solve the Einstein field equations, which yield the modified Friedmann equation:
\begin{eqnarray}
H^2 = \frac{8 \pi G }{3} \rho + \sigma^2, 
\label{H}
\end{eqnarray}
where $H$ is the cosmic expansion rate, $G$ is the Newtonian gravitational constant, $\rho$ is the energy density, and $\sigma$ is the shear stress. In the case of temperature fluctuations within the quadrupole expansion (assuming $T_q$ is sufficiently small and expanding up to second order), the energy density in the radiation-dominated era takes the modified form:
\begin{eqnarray}
\rho \simeq g^* A T^4
\left( 1+ \frac{8}{15} T_{ab}T^{ab} \right),
\label{rho}
\end{eqnarray}
where $A = \pi^2 / 30$, and $g^*$ denotes the effective degrees of freedom for relativistic species.

In Eq.\,(\ref{H}), $\sigma$ is time-dependent, and its evolution is governed by the following equation derived from the Einstein equation:
\begin{eqnarray}
    \dot{\sigma} + 3H \sigma = \pi_{11}.
\label{sigma_ev}    
\end{eqnarray}
The $\pi_{11}$ is determined from the general anisotropic stress tensor of relativistic particles, $\pi_{ab}$:
\begin{eqnarray}
    \pi_{ab} \equiv \int \mathrm{d}\Omega \, e_{\langle 
 a} e_{b \rangle} \int \mathrm{d}E \, E^3 f(E,T_\mathrm{tot}),
\label{pidef}
\end{eqnarray}
where $\mathrm{d}\Omega$ is the solid angle, and $e_{\langle a} e_{b \rangle} \equiv e_a e_b - \delta_{ab} / 3$ represents the traceless part of $e_a e_b$. As assumed, at second order in $T_{ab}$, one finds:
\begin{eqnarray}
    \pi_{ab} = \frac{8}{15} A g^* T^4 T_{ab}.
\label{pi}
\end{eqnarray}

Another important condition is the conservation of energy-momentum, $\partial_\mu {\cal T}^{\mu \nu} = 0$, which leads to the following continuity equation:
\begin{eqnarray}
    \dot{\rho} + 4 H \rho + \sigma_{ab} \pi^{ab} = 0.
\label{CE}
\end{eqnarray}
From the continuity equation, the time-temperature relation is given as 
\begin{eqnarray}
    \frac{dT}{dt} = - \left[ H + \frac{4 \sigma T_q}{5 +16 T_q^2} \right] T.
\label{dTdt}  
\end{eqnarray}
We note that Eq.\,(\ref{dTdt}) reduces to the standard formula, $dT/dt = - H T$, when $\sigma=0$ or $T_q=0$. Then, using Eq.\,(\ref{dTdt}), Eq.\,(\ref{sigma_ev}) can be rewritten in terms of the temperature evolution as:
\begin{eqnarray}
    \frac{d \sigma}{dT} = \left( \frac{8}{15} A g^* T^4 T_q - 3 H \sigma \right) \left( \frac{dT}{dt} \right)^{-1}.
\label{sigma_dT}    
\end{eqnarray}

\begin{figure}[t]
    \centering
    \includegraphics[width=0.85\linewidth]{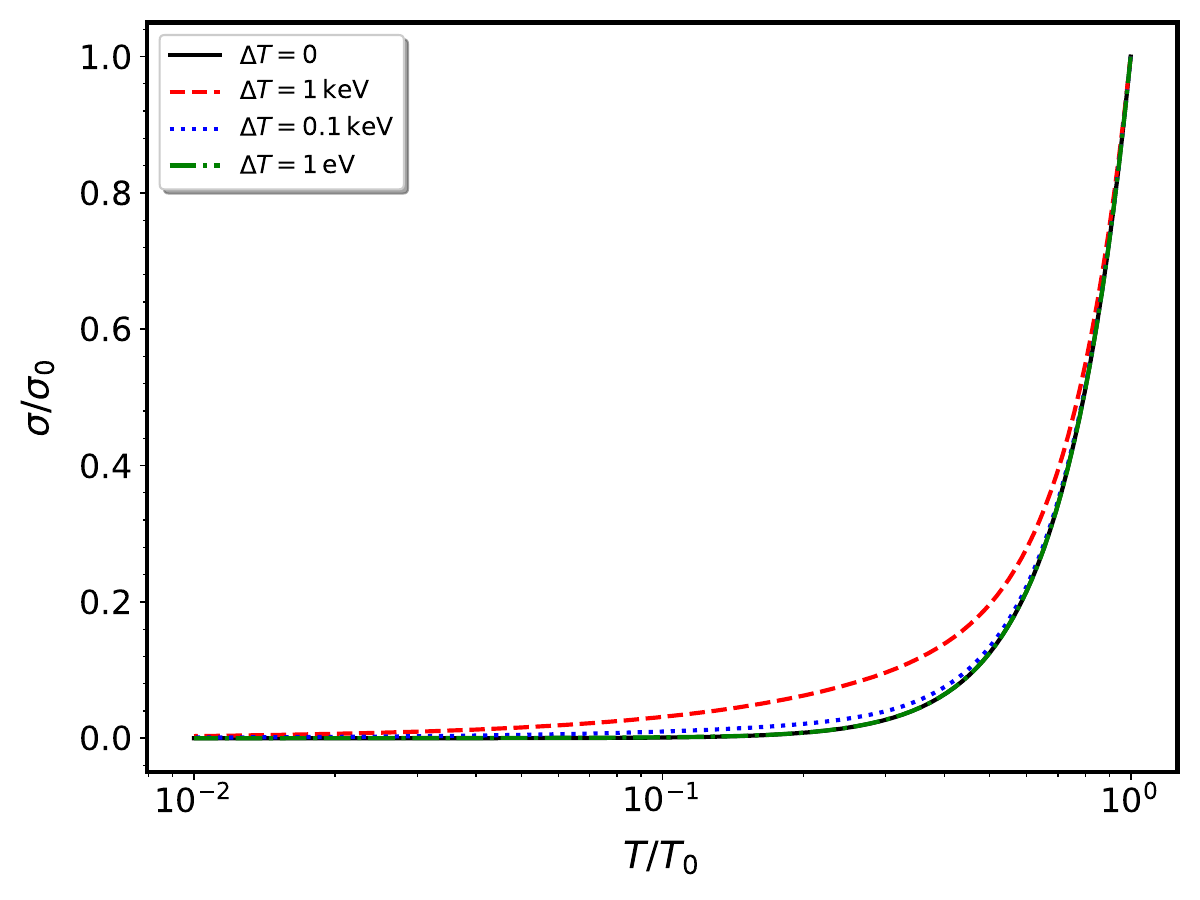}
    \caption{Numerical solution of $\sigma/\sigma_0$ as a function of $T/T_0$. The black solid, red dashed, blue dotted, and green dash-dotted lines represent the solutions for $\Delta T = 0$, $\Delta T = 1\,{\rm keV}$, $\Delta T = 0.1\,{\rm keV}$, and $\Delta T = 1\,{\rm eV}$, respectively.}
    \label{fig:sigma}
\end{figure}

Figure \ref{fig:sigma} shows the evolution of $\sigma$ as a function of $T$. As shown in the figure, for small $\Delta T$, $\sigma$ approaches the solution for the case of $\Delta T = 0$. Notably, the solution for $\Delta T = 0$ can be obtained analytically as:
\begin{eqnarray}
    \sigma = \sigma_0 \left( \frac{T}{T_0} \right)^3,
\label{sigma_sol}    
\end{eqnarray}
where $\sigma_0$ and $T_0$ are initial value of $\sigma$ and initial temperature, respectively.

In principle, the temperature fluctuation parameter $T_q$ should be included in the cosmic expansion rate along with shear stress in BBN calculations. However, cosmic microwave background (CMB) observations indicate that the temperature fluctuation is approximately $\Delta T \sim \mathcal{O}(10^{-5}\,{\rm eV})$. Since the BBN temperature scale is $T_{\rm BBN} \sim \mathcal{O}({\rm MeV})$, the corresponding value of $T_q$ in the BBN epoch is estimated to be $T_q \sim \mathcal{O}(10^{-11})$. Consequently, we can neglect $T_q$, allowing Eq.\,(\ref{H}) to be rewritten as follows:
\begin{eqnarray}
    H^2 = \frac{8 \pi G}{3} \rho + \sigma_0^2 \left( \frac{T}{T_0} \right)^6,
\label{H_sigma}    
\end{eqnarray}
here the correction term of the $\rho$ in Eq.\,(\ref{rho}) can also be neglected as it is proportional to $T_q^2 = (\Delta T/T)^2 \sim {\cal O}(10^{-22})$ at the $T_{\rm BBN} \sim 1\,{\rm MeV}$. 

%%%%%%%%%%%%%%%%%%%%%%%%%%%%%%%%%%%%%%%%%%%%%%%%%%%%%%%%%%%%%%%%%%%%%%%%%%%%
%%%%%%%%%%%%%%%%%%%%%%%%%%%%%%%%%%%%%%%%%%%%%%%%%%%%%%%%%%%%%%%%%%%%%%%%%%%%
\section{Constraints from BBN}
\label{bbn}
To incorporate the modified $H$ in Eq.\,(\ref{H_sigma}) into the BBN calculation, we employ the BBN calculation code \cite{Kawano:1992ua, Smith:1992yy}, implementing updated reaction rates from the JINA REACLIB database \cite{Cyburt:2010} and improved reaction rates for deuterium (D) \cite{Coc:2015bhi}. As input parameters, we adopt the central value of the neutron mean lifetime provided by the Particle Data Group, $\tau_n = 878.6 \pm 0.6\,s$ \cite{ParticleDataGroup:2022pth}, and the lower limit of the baryon-to-photon ratio, $\eta = (6.104 \pm 0.058) \times 10^{-10}$, which corresponds to the baryon density based on the $\Lambda$CDM model (TT, TE, EE+lowE) from Planck observations of the cosmic microwave background, $\Omega_bh^2 = 0.02230 \pm 0.0021$ \cite{Planck:2018vyg}. The initial temperature is set to $T_0 = 0.8617\,{\rm MeV} (=10^{10}\,{\rm K})$.

In the radiation-dominated era, where $\rho \propto T^4$, the second term in Eq.\,(\ref{H_sigma}) is significant near the initial temperature region. Subsequently, as the universe cools, this term rapidly decays. Consequently, the correction term arising from shear stress primarily affects the early stages of BBN.

\begin{figure}[t]
    \centering
    \includegraphics[width=0.85\linewidth]{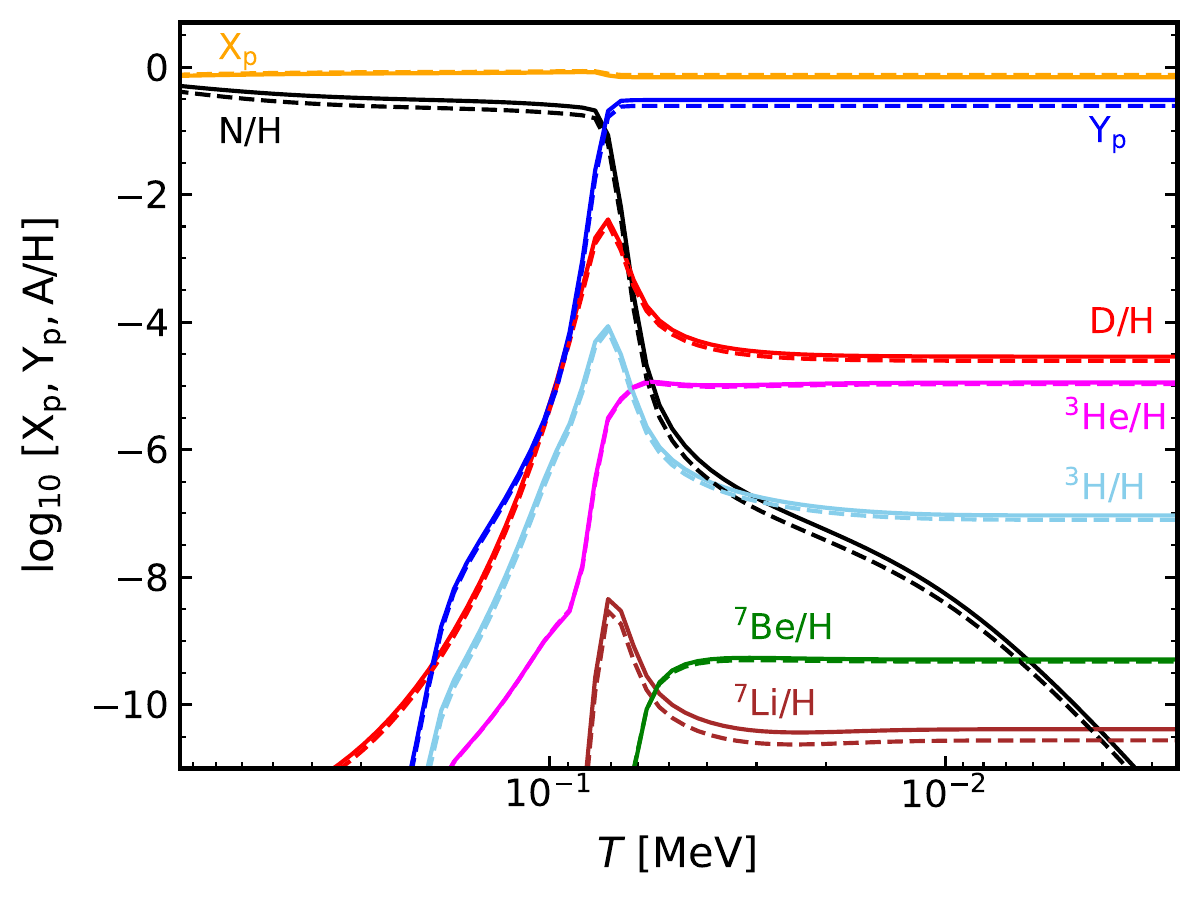}
    \caption{Evolution of primordial abundances as a function of temperature. Solid and dashed lines indicate the results with $\sigma_0= 5.0 \times 10^{-13}\,{\rm eV}$ and $\sigma_0=0$ (standard), respectively. ${\rm X_p}$ denotes the mass fraction of proton, ${\rm Y_p}$ the mass fraction of $^4{\rm He}$, and ${\rm A/H}$ the abundances of element A labeled in the figure. ($N$ stands for the neutron.) }
    \label{fig:ab_evol}
\end{figure}

Figure \ref{fig:ab_evol} illustrates the evolution of primordial abundances as a function of the BBN temperature for both the standard Friedmann equation and the case with $\sigma_0 = 5.0 \times 10^{-13}\,{\rm eV}$. As discussed above, the presence of $\sigma_0$ increases the Hubble expansion rate $H$ beyond that of standard cosmic expansion. Consequently, neutron freeze-out occurs earlier, resulting in an increased neutron abundance, as shown in Fig.\,\ref{fig:ab_evol}.

The higher neutron abundance enhances the reaction $^1{\rm H}(n,\gamma)^2{\rm H}$, thereby increasing the D abundance. The increased D abundance, in turn, enhances the rates of subsequent nuclear reactions, including ${\rm D}(d,p)^3{\rm H}$, ${\rm D}(d,n)^3{\rm He}$, $^3{\rm H}(d,n)^4{\rm He}$, and $^3{\rm He}(d,p)^4{\rm He}$. Consequently, as $\sigma_0$ increases, the abundances of ${\rm ^3He}$ and ${\rm ^4He}$ also increase. Furthermore, the increased abundances of $^{3}{\rm H}$ and $^{3}{\rm He}$ enhance the reactions $^3{\rm H}(\alpha, \gamma)^7{\rm Li}$ and $^{3}{\rm He}(\alpha, \gamma)^7{\rm Be}$, leading to an increase in $^7{\rm Li}$ and $^7{\rm Be}$. As a result, the abundance of $^7{\rm Li}$ rises with increasing $\sigma_0$. In \cite{Jang:2024jso}, the deviation from standard evolution equation is studied for the energy-momentum-squared gravity model.  Energy-momentum-squared gravity model's correction term of $H$ scales proportionally to $T^8$. The  primordial abundances as a function of $\sigma_0$ in a similar way as it has in \cite{Jang:2024jso}, providing a model independent robustness of the BBN test.

\begin{figure}[t]
    \centering
    \includegraphics[width=0.85\linewidth]{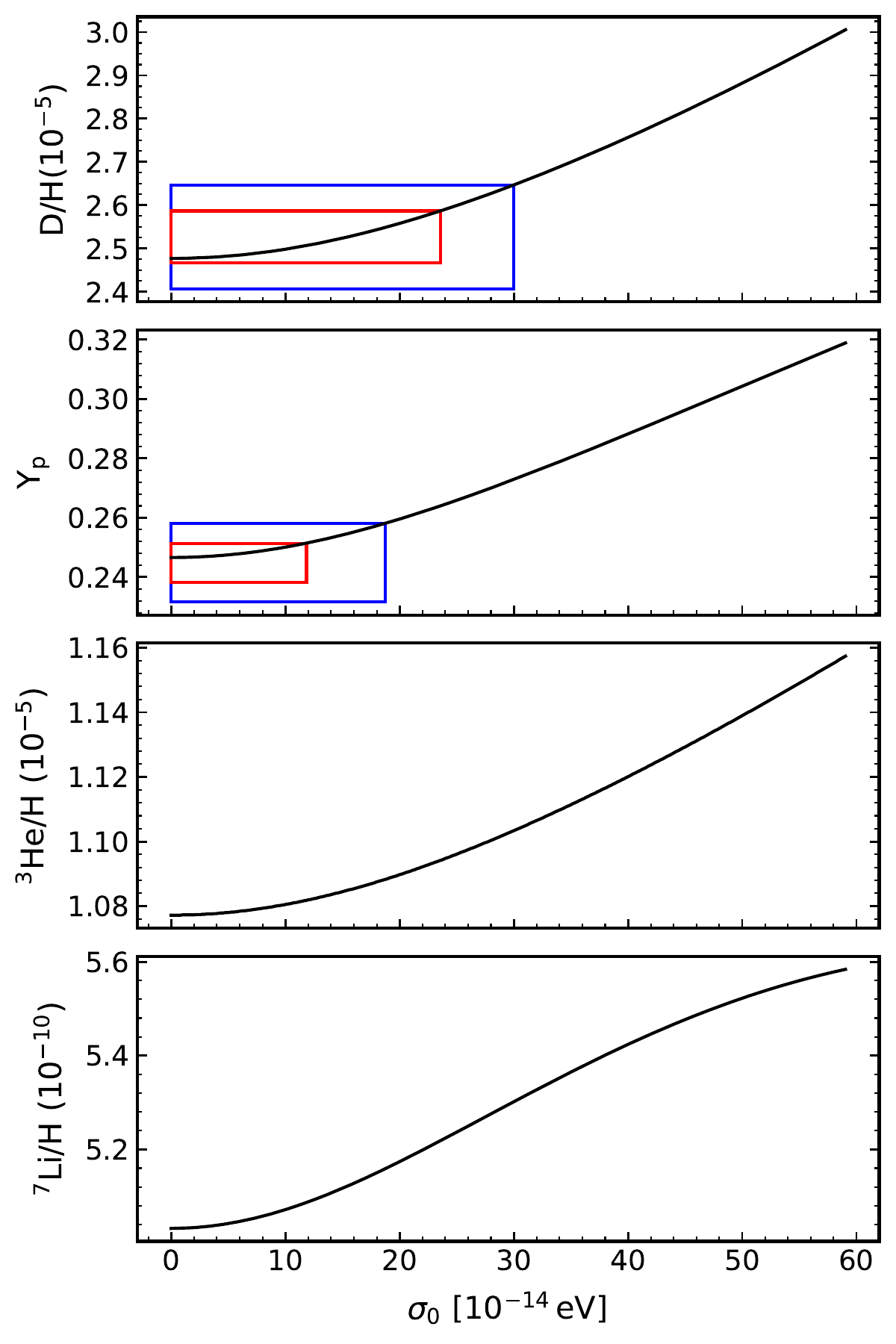}
    \caption{Final abundances of of D, $^4{\rm He}$, $^3{\rm He}$, and $^7{\rm Li}$ as a function of $\sigma_0$. In the first and second panels, the red and blue boxes indicate constrained regions by the observational data within $2\sigma$ and $4 \sigma$ range, respectively. }
    \label{fig:final_abund}
\end{figure}

Figure \ref{fig:final_abund} presents the final abundances as a function of $\sigma_0$, reflecting the trends discussed above. For D/H, compared to the metal-poor Lyman-$\alpha$ absorption measurement ${\rm D/H} = \left(2.527 \pm 0.030 \right) \times 10^{-5}$ \cite{Cooke:2017cwo}, we find that the upper limit of $\sigma_0$ is constrained to $2.36 \times 10^{-13} \,{\rm eV}$ at $2\sigma$ and $3.00 \times 10^{-13}\,{\rm eV}$ at $4\sigma$. 

For the mass fraction of $^4{\rm He}$, ${\rm Y}_p$, based on observational data from metal-poor extragalactic H II regions, ${\rm Y}_p = 0.2448 \pm 0.0033$ \cite{Aver:2022}, we find that the upper limit of $\sigma_0$ is constrained to $1.18 \times 10^{-13}\,{\rm eV}$ at $2\sigma$ and $1.88 \times 10^{-13}\,{\rm eV}$ at $4\sigma$. Since the constraint on ${\rm Y}_p$ is more stringent than that on ${\rm D/H}$, the BBN constraints primarily follow the limit set by ${\rm Y}_p$.

For ${\rm ^3He}$ in the third panel of Fig.\,\ref{fig:final_abund}, the abundance is consistent with the observational upper limit of $^3{\rm He/H} = \left(1.1 \pm 0.2\right) \times 10^{-5}$ \cite{Bania:2002yj}. On the other hand, in the fourth panel, the abundance of ${\rm ^7Li}$ increases as $\sigma_0$ increases. Consequently, for all $\sigma_0$, the abundance of ${\rm ^7Li}$ exceeds the observational value of $^7{\rm Li/H} = \left(1.58 \pm 0.31\right) \times 10^{-10}$ \cite{Sbordone:2010}, indicating that this anisotropic model cannot resolve the primordial lithium problem.

%%%%%%%%%%%%%%%%%%%%%%%%%%%%%%%%%%%%%%%%%%%%%%%%
%%%%%%%%%%%%%%%%%%%%%%%%%%%%%%%%%%%%%%%%%%%%

\section{Conclusion:}
\label{conc}
In this study, we explore the effects of shear stress and temperature fluctuations on BBN within the B-I anisotropic spacetime. By incorporating these effects into the modified Friedmann equation, we present the effects of the shear stress term on the abundances of primordial abundances and derive observational constraints on the shear stress parameter $\sigma_0$.

We find that the temperature quadrupole parameter $T_q$ is constrained by CMB observations and can be safely neglected in BBN calculations. On the other hand, the shear stress term modifies the cosmic expansion rate during the BBN epoch, being proportional to $T^6$. From BBN observations, we constrain $\sigma_0 \leq 1.18 \times 10^{-13}\,{\rm eV}$ within $2\sigma$ and $\sigma_0 \leq 1.88 \times 10^{-13}\,{\rm eV}$ within $4\sigma$. Of course, $\sigma_0 =0$ gives back the standard isotropic scenario. These results suggest that primordial nucleosynthesis can serve as a sensitive probe of early-universe anisotropies, providing constraints on deviations from standard cosmology.

%%%%%%%%%%%%%%%%%%%%%%%%%%%%%%%%%%%%%%%%%%%%%%%%%%%%%%%%%%%%%%%%%%%%%%%%%%%%%%%%%%%%%%%%%%%%%%%%%%%%%%%%%%%%%
%%%%%%
\section*{Acknowledgments:}
The authors would like to thank M. Sami for fruitful discussions.

JWP and MKC are supported by the Basic Science Research Program through the National Research Foundation of Korea, funded by the Ministry of Education, Science, and Technology (Grant No. NRF-2017R1D1A1B06032249, Grant No. NRF-2021R1A6A1A03043957).

Work of MRG is supported by the Science and Engineering Research Board(SERB), DST, Government of India under the Grant Agreement number CRG/2022/004120(Core Research Grant).  

%\bibliographystyle{utphys}
%\bibliography{wipbh}

%%%%%%%%%%
%%%%%%%%%%%%

%%%%%%%%%%%%%%%%%%%%%%%%%%%%%%%%%%%%%%%%%%%%%%%%%%%%%%%%%%%%%%%%%%%%%%%%%%%%%%%

\clearpage
\appendix
%\onecolumngrid
%\section*{Supplemental Material}

%%%%%%%%%%%%%%%%%%%%%%%%%%%%%%%%%%%%%%%%%%%%%%%%%%%%%%%%%%%%%%%%%%%%%%%%%%%%%%%
\end{document}